\documentclass[prb,amsmath,superscriptaddress,twocolumn,showpacs]{revtex4-1}
\usepackage{bm}
\usepackage{amssymb}
\usepackage{colordvi}
\usepackage{graphicx}
\usepackage{hyperref}

\newcommand{\ov}[1]{\overline{{#1}}}
\newcommand{\be}{\begin{equation}}
\newcommand{\ee}{\end{equation}}
\newcommand{\bea}{\begin{eqnarray}}
\newcommand{\eea}{\end{eqnarray}}

\begin{document}

\title{Thermoelectric three-terminal hopping transport through 
  one-dimensional nanosystems}

\author{Jian-Hua Jiang}
\affiliation{Department of Condensed Matter Physics, Weizmann Institute of
  Science, Rehovot 76100, Israel}
\author{Ora Entin-Wohlman}
\affiliation{Department of Physics and the Ilse Katz Center for Meso-
  and Nano-Scale Science and Technology, Ben Gurion University, Beer
  Sheva 84105, Israel}
\author{Yoseph Imry}
\affiliation{Department of Condensed Matter Physics, Weizmann Institute of
  Science, Rehovot 76100, Israel}

\date{\today}

\begin{abstract}
A  two-site nanostructure (e.g, a ``molecule" ) bridging two
conducting leads and connected to a phonon bath is considered. The two
relevant  levels closest to   the  Fermi  energy are connected each to
its lead. The leads  have slightly different temperatures and chemical
potentials and the nanostructure is also coupled to a thermal
(third)  phonon bath. The $3\times 3$ linear transport (``Onsager")
matrix is evaluated, along with the ensuing new figure of merit, and
found to be very favorable for thermoelectric energy conversion. 
\end{abstract}


\maketitle

\vspace{-1cm}

\section{Introduction}
In thermoelectric transport temperature differences can be converted
to (or generated by) electric voltages. Such phenomena have already
found several useful applications. Current research is motivated  by
the need for higher performance thermoelectrics as well as the pursuit
of understanding of various relevant microscopic processes (especially the
inelastic ones). Theory  \cite{CUTLER,Joe,interference} predicts that high values
of the  thermopower follow when the carriers' conductivity  depends
strongly on energy. Indeed,  in bulk systems, the thermoelectric
effects  necessitate electron-hole asymmetry, which is often rather
small. However, in nanosystems,  
such asymmetry can arise in individual samples in ensembles with
electron-hole symmetry on average. Moreover,  inelastic processes
and interference effects may play nontrivial roles in
thermoelectric transport.\cite{interference}
It is known that the thermoelectric performance is governed by the
dimensionless figure of merit  $ZT$,\cite{Honig} where $T$ is the
common temperature of the system and $Z=\sigma
S^{2}/(\kappa_{e}+\kappa_{ph})$, with $\sigma$ being the electrical
conductivity, $S$  the  Seebeck coefficient, and $\kappa_{e}$ and
$\kappa_{ph}$   the electronic and the phononic heat conductivities,
respectively. Both  $\kappa_{e}$ and $\kappa_{ph}$  can be smaller in
nanosystems  \cite{Dresselhaus} than in bulk ones, opening a route for
better thermoelectrics. 

Mahan and Sofo \cite{mahan} have argued that the best
thermoelectric efficiency can be achieved in systems where i) the
energy width of the main conducting channel is very narrow,  and ii)
the phonon thermal conductivity is as small as possible.  It was
suggested that ii) can be also realized in nanoscale composite
structures where phonons are scattered by large variations in geometry
and abundant interfaces,\cite{inter1,interface} while  i) leads
\cite{mahan} to a very small $\kappa_{e}/(S^{2}\sigma)$,  as indeed
has been confirmed in studies of  quantum dot arrays. \cite{Linke}

Here we consider three-terminal thermoelectric transport in small
one-dimensional (1D) nanosystems  accomplished  via inelastic
phonon-assisted hopping, and show that such processes lead to several
nontrivial properties. Although thermal transport properties of
mesoscopic structures have been studied  in the
past, \cite{Joe,interference,meso-rmp,molecule} investigations 
of three-terminal thermoelectric transport are just at their
infancy. \cite{3t} Our main conclusions are drawn from the  simple,
but important, two localized-state  junction in which hopping is
{\em nearest-neighbor}.\cite{Oren} Later, we briefly discuss larger 1D
systems, which exhibit rather surprising features of the thermoelectric
transport. We show that such systems  may have a high figure of merit,
as $\kappa_{e}$  can become extremely small, while the thermopower, $S$, remains finite.  Some of the
thermoelectric transport  coefficients we find  correspond to
transferring electric/thermal current via temperature difference
between the electron system and a suitable phonon bath. Hopefully,
such systems can be achieved within current technology and  be useful
in applications.

There are several related ideas in the literature.
Reference~\onlinecite{Edwards} presented an early, ingenious, way to
cool a finite 2D electron gas (which plays the role of the thermal
bath) at low temperatures  by elastic electron transitions to/from the
leads. All the energies involved are only of order $k_{\rm B}
T$. Reference~\onlinecite{Prance} provided an experimental realization of
some of the suggestions of Ref.~\onlinecite{Edwards},  with further
analysis.  Reference~\onlinecite{Ratchett} demonstrated  a quantum
ratchet, converting the nonequilibrium noise of a nearby quantum
point contact to dc current. Reference~\onlinecite{Zippilli} suggested a
sophisticated Carbon nanotube  structure, designed to extract energy
from a discrete local oscillator at ultralow temperatures. The present
work considers the full three-terminal case, where the energies involved
can be larger than $k_{\rm B} T$, and a real reservoir can be cooled,
not just one or several degrees of freedom.

\section{Model system}

\subsection{Hamiltonian}
The Hamiltonian, $H=H_e+H_{e-ph}+H_{ph}$, consists of the electronic
and phononic parts and the electron-phonon interaction. The
electronic part is (electronic operators are denoted by $c$ and
$c^{\dagger}$) 
\begin{align}
&H^{}_{e}=\sum_{i}E^{}_{i}c^{\dagger}_{i}c^{}_{i}+\sum_{k(p)}\epsilon^{}_{k(p)}c^{\dagger}_{k(p)}c^{}_{k(p)}\nonumber\\
&+\Bigl (\sum_{i,k(p)}J^{}_{i,k(p)}c^{\dagger}_{i}c^{}_{k(p)}+\sum_{i}J^{}_{i,i+1}c^{\dagger}_{i}c^{}_{i+1}+{\rm H.c.}\Bigr )\ .
\end{align}
Here $i$ labels the localized states, of energies $E_{i}$ (including usual Coulomb-blockade effects; i.e., it is assumed implicitly that a large Hubbard interaction confines the occupation of each level to be 0 or 1)  and $k$ ($p$) marks the 
extended states in the left (right) lead, of energies $\epsilon_{k}$ ($\epsilon_{p}$) (all energies are measured 
from the common chemical potential). The matrix element 
coupling the localized states to each other is $J_{i,j}$, and those 
coupling them to the lead states are $J_{i,k(p)}$. 
All are exponentially decaying, with a  localization length $\xi$, e.g., 
\begin{align}
J^{}_{i,k(p)}=\alpha^{}_{e}\exp\Bigl (-\frac{|x^{}_{i}-x^{}_{L(R)}|}{\xi}\Bigr )\ ,
\end{align}
with $x_i$ and
$x_{L(R)}$ being the coordinates of the center of the localized states and
the left (right) boundary, and the prefactor $\alpha_{e}$ yielding the coupling energy. The electron-phonon interaction 
is
\begin{align}
H^{}_{e-ph}=\sum_{\bf q}M^{}_{{\bf q},ij}c^{\dagger}_{i}c^{}_{j}(a^{}_{\bf q}+a^{\dagger}_{-{\bf q}})+{\rm H.c.}\ ,\label{EPH}
\end{align}
where the phonon modes, of wave vector  ${\bf q}$ and frequency
$\omega_{\bf q}$,  are described by the operators $a^{\dagger}_{\bf
  q}$, $a^{}_{\bf q}$. Their Hamiltonian is $H_{ph}=\sum_{\bf
  q}\omega_{\bf q}a^{\dagger}_{\bf q}a^{}_{\bf q}$ (we use units where $\hbar =1$).
The electron-phonon coupling
is $M_{{\bf q},ij}=\alpha_{e-ph}\exp(-|x_{i}-x_{j}|/\xi )$, with
$\alpha_{e-ph}$ being the electron-phonon coupling energy.
The transport through the system is governed by hopping when the
temperature is above a crossover temperature,  $T_x$, estimated below in Sec. \ref{TS}
for the most important two-site case. At lower temperatures the
dominant transport is via tunneling.  The two-site example of our
system is depicted in the upper panel of  Fig. \ref{fig1}.

\subsection{Hopping and interface resistors}

\begin{figure}[htb]
\includegraphics[height=3.8cm]{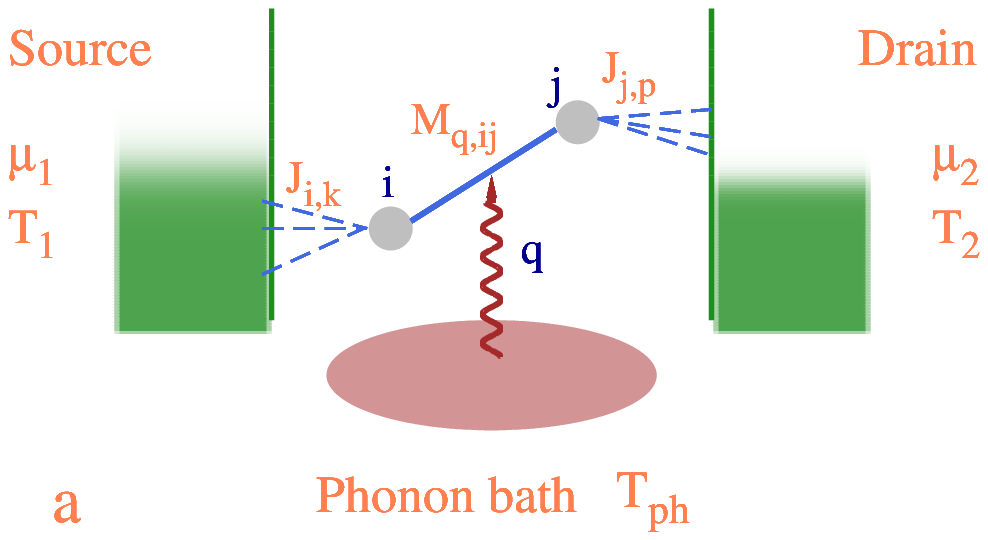}
\includegraphics[height=3.8cm]{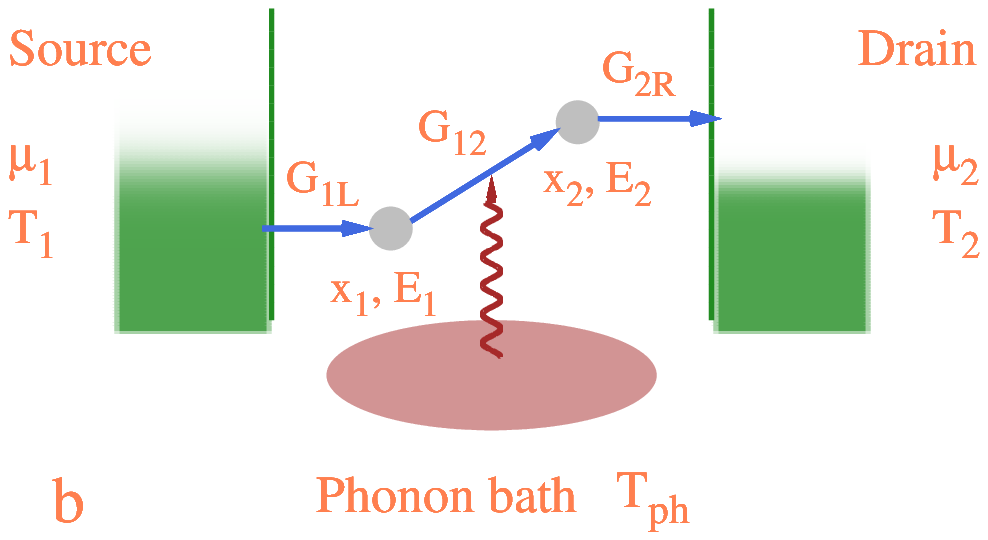}
\caption{(Color online) a. A two localized-state ($i$ and $j$, gray
  points) system coupled to two leads, of temperatures $T_{L}$ and
  $T_{R}$, and chemical potentials $\mu_{L}$ and $\mu_{R}$ (with the
  choice $\mu_L>\mu_R$ and $T_L>T_R$).  The phonon bath temperature
  is  $T_{ph}$. The localized states are coupled (dotted lines)  to
  the  continuum of states in the leads, and are also coupled (the
  wavy line) to the phonon bath;  b. The effective resistors
  representing the system: The straight (blue)   arrows indicate the
  net electronic currents  and the wavy (brown) one the phonon heat
  current, with  $G_{1L}$, $G_{2R}$, and $G_{12}$ being the
  conductances of the tunneling  and the hopping resistors,
  respectively.}
\label{fig1}
\end{figure}

The system described above bridges two electronic leads, held at 
slightly different temperatures and chemical potentials, $T_{L}$, $\mu_{L}$,  
and $T_{R}$, $\mu_{R}$, 
such that the common temperature is
$T\equiv (T_{L}+T_{R})/2$.  The golden-rule transition rate $\Gamma_{ij}$, between
two localized states, located at $x_{i}$ and $x_{j}$ and having
energies $E_{i}<0<E_{j}$,\cite{MA}  necessitates the inelastic
electron-phonon scattering (\ref{EPH}), and reads 
\begin{align}
\Gamma^{}_{ij}=2\pi \Gamma^{}_{in}
f^{}_{i}(1-f^{}_{j})N^{}_{B}(E^{}_{ji})\ ,\label{GAMA}
\end{align}
where $E_{ji}\equiv E_{j}-E_{i}$, the carriers' local Fermi function is
\begin{align}
f^{}_{i}= \Bigl [\exp\Bigl (\frac{E^{}_{i}-\mu^{}_{i}}{k^{}_{\rm B}T_{i}^{}}\Bigr )+1 \Bigr ]^{-1}_{}\ ,
\end{align}
and $N_{B}$ is the Bose function
\begin{align}
N^{}_{B}=\Bigl [\exp\Bigl (\frac{\omega^{}_{\bf q}}{k^{}_{\rm B}T^{}_{ph}}\Bigr )-1\Bigr  ]^{-1}_{}\ ,
\end{align}
determined by $T_{ph}$,   the temperature of the local phonon
bath (see Fig. \ref{fig1}). We assume that this  phonon bath is strongly coupled to
a thermal reservoir and is thermally isolated as much as possible from the 
leads, such that its temperature is determined by that
reservoir. On the other hand, phonons in the leads are in good
thermal contact with the electrons there and  share the same
temperature. These assumptions are further elaborated upon in Sec. \ref{3tf}. In Eq. (\ref{GAMA}), $\Gamma_{in}=|M_{{\bf q},ij}|^{2}\nu_{ph}
(|E_{ij}^{}|)$,  where $\nu_{ph}$ is the phonon density of states.
The linear hopping conductance at long distances 
($|x_{i}-x_{j}|\equiv |x_{ij}|\gg \xi$) and high energies  ($|E_{i}|$, $|E_{j}|\gg k_{\rm B}T$) of
the bond  $ij$   is  \cite{MA}
\begin{align}
&G^{}_{ij}\sim  \frac{e^{2}}{k_{\rm B}T} |\alpha^{}_{e-ph}|^{2}\nu_{ph}^{}
(|E_{ij}^{}|) \eta^{-1}_{ij} \ ,\nonumber\\
&\eta^{}_{ij}=\exp\Bigl (\frac{2|x_{ij}^{}|}{\xi}\Bigr )\exp\Bigl (\frac{|E^{}_{i}|+
|E^{}_{j}|+|E^{}_{ij}|}{2k^{}_{\rm B}T}\Bigr )\ .\label{HOP}
\end{align}

As opposed to Eq. (\ref{GAMA}), the tunneling conduction from, say,
site $i$ to  the left lead can be accomplished by
elastic tunneling processes with a transition rate $\Gamma_{iL}
= \gamma_{iL} f_i[1-f_L(E_i)]$,  where $\gamma_{iL}=2\pi |J_{ik}|^2
\nu_L(E_i)$ and  $f_{L}$ and  $\nu_L$ are the Fermi distribution and density of states of the left
lead. The corresponding linear interface conductance is then
$G_{iL}\simeq e^{2}|\alpha_{e}|^{2}\nu_{L}(E_{i})(k_{\rm B}T)^{-1}\exp[-2|x_{iL}|/\xi-|E_i| /(k_{\rm B}T)]$.
This conductance (and the interface conductance at the right lead) will be assumed to 
be much larger than the hopping conductance between the two localized
states.

\subsection{The two-site case}

\label{TS}
The thermopower in the hopping regime has been discussed by
Zvyagin.\cite{ZO} The simplest example is that of a  two-site system
($i,j=1,2$) depicted in Fig. \ref{fig1}, which describes, e.g. a
diatomic molecule \cite{Oren} or a series-connected double quantum
dot.\cite{Linke} In such a case transport is accomplished by nearest-neighbor  
hopping. As site 1 (2) is in a  good contact with left (right)
lead, we may assume that the local chemical potential and temperature
there are  $\mu_{L(R)}$  and  $T_{L(R)}$. The transport is dominated
by the hopping from 1 to 2 when the temperature is higher than
$T_{x}$. This temperature is estimated from the requirement that the
elastic tunneling conductance  across the system, $G_{tun}$, is
comparable to the hopping one. The former is given by
the transmission $\sum_{i=1,2}\Gamma_{iL}(E)\Gamma_{iR}(E)/[(E-E_{i})^{2}+(
\Gamma_{iL}(E)+\Gamma_{iR}(E))^{2}/4]$, 
where
the tunneling rates are 
$\Gamma_{iL(R)}(E)=2\pi |J_{i,k(p)}|^2\nu_{L(R)}(E)$. Since site $1$
(2) is coupled mostly to the left (right) lead,
we use their perturbation-theory mixtures,
governed by
the small parameter $J_{12}/E_{21}$. At low
temperatures and  for $|E_1|,|E_2|\gg \Gamma_{1},
\Gamma_{2}$, where $\Gamma_{i}\equiv\Gamma_{iL}+\Gamma_{iR}$.,  we
estimate
\begin{align}
G^{}_{tun}\sim
e^{2}E^{-2}_{1}|\alpha_{e}|^{6}\nu^{}_{L}(0)\nu_{R}^{}(0)E_{2}^{-2}\exp\Bigl
(-\frac{2W}{\xi}\Bigr )\ , 
\label{Gtun}
\end{align}
where $W$ is  the system length between the leads.
The hopping conductance is given by Eqs. (\ref{HOP}) (with $i,j=1,2$).
Comparing those two, with exponential accuracy, the elastic tunneling
mechanism can be important \cite{SL} only when 
\begin{align}
&\exp\Bigl [-\frac{|E_1|+|E_2|+|E_{12}|}{2k_{\rm B}T}\Bigr ]\ll
\exp\Bigl [-2\frac{W-|x_{12}|}{\xi}\Bigr ]  \nonumber\\
&\ \ \rightarrow \ \ \eta_{12}\gg \exp (2W/\xi)\ ,
\end{align}
giving $k_{\rm B}T_x\sim (|E_1|+|E_2|+|E_{12}|)\xi/[4(W-|x_{12}|)]$.

\section{Three-terminal thermoelectric linear transport} 
\subsection{Transport equations}
The electronic particle current through the system is, as in Eq. (\ref{GAMA})  allowing for the temperature and chemical potential differences,
\begin{align}
I^{}_{N}=\Gamma^{}_{12}-\Gamma^{}_{21}\ .\label{IN}
\end{align}
 For an electron transferred from left to right, the bath gives an energy $-E_{1}$
($E_{2}$) to the left (right) lead, and thus the phonons transfer the energy $E_{21}$ to the electrons. 
A net energy of  $\overline{E}\equiv [E_{1}+E_{2}]/2$ is transferred from left to right. Hence, the net electronic energy current,  $I_Q^e$,
and the heat current \cite{Heat} exchanged between the electrons and the phonons,  $I_Q^{pe}$, are
\begin{align}
I_Q^e =\overline{E}I^{}_N, \ \ \ {\rm and}\ \ \  I_Q^{pe} = E^{}_{21}I^{}_N \ .\label{IQ}
\end{align}
The linear-response 
transport coefficients are obtained by expanding Eqs. (\ref{IN}) and (\ref{IQ}) to first order in $\delta T \equiv T_L-T_R$,
$\delta\mu \equiv \mu_L-\mu_R$, and $\Delta T \equiv T_{ph}-T$,  
\begin{align}
\left( \begin{array}{c}
I^{}_e\\ I_Q^e\\ I_Q^{pe}\end{array}\right) =
\left( \begin{array}{cccc}
G & L^{}_1 & L^{}_2 \\
L^{}_1 & K_e^0 & L^{}_{3} \\
L^{}_2 & L^{}_{3} & K^{}_{pe} \\
\end{array}\right) \left(\begin{array}{c}
\delta\mu/e\\ \delta T/T\\ \Delta T/T \end{array}\right) \ ,\label{FIRO}
\end{align}
where $I_e=eI_N$ is the charge current.
All transport coefficients in Eq. (\ref{FIRO}) are given in terms of the linear
hopping conductance $G$ [given by Eqs. (\ref{HOP})], 
\begin{align} 
&L^{}_{1}=\frac{G}{e}\ov{E}\ ,\     L_2^{} = \frac{G}{e}  E^{}_{21}\ , \nonumber\\
& K_e^0 = \frac{G}{e^2}\ov{E}^2\ ,\  
L_{3}^{} = \frac{G}{e^2}\ov{E}E^{}_{21}\ ,\ K_{pe}=\frac{G}{e^2}E_{21}^2 \ .\label{TC}
\end{align}
Note that $L_2$, $L_{3}$, and $K_{pe}$ are related to
$E_{21}$, and  $I_Q^{pe}$ vanishes linearly  with the latter.

The transport coefficients $L_2$  ($L_{3}$) correspond to, e.g.,  generating {\em
electronic current} ({\em energy current}) via the temperature
difference $\Delta T$.\cite{3t} 
When reversed, this process performs as  a refrigerator: Electric current
pumps heat current away from the phononic system and {\em cools it down}. In analogy
with the usual two-terminal thermopower $S$, here we use the  three-terminal thermopower \cite{3t} of this process,
\begin{align}
S^{}_p = \frac{L^{}_2}{TG} = \frac{k^{}_{\rm B}}{e}\frac{E^{}_{21}}{k^{}_{\rm B}T} \   .\label{SP}
\end{align}
Note that $S_p$ of our  model  can be very large  as the energy taken from the phonons 
per transferred electron can be several
times  $k_{\rm B}T$.


\subsection{Two-terminal  figure of merit, for $\Delta T=0$} 
A significant feature of our setup  is that the electronic heat
conductance can vanish while the thermopower stays finite
\begin{align}
K_e^{} = K_e^0 - \frac{L_1^2}{G} = 0 \ ,\ \ 
S=\frac{L^{}_1}{TG}=\frac{k_{\rm B}}{e}\frac{\ov{E}}{k_{\rm B}T} \ .
\end{align}
According to Ref.~\onlinecite{mahan}, the largest two-terminal figure of merit is achieved
in systems with the smallest $\kappa_e/\sigma S^2$ (provided that $S$ stays finite). Here this ratio vanishes, and then $Z $
is
limited by  $\kappa_{ph}$. The latter can be minute in
nanosystems.\cite{Dresselhaus} Moreover,  
it can be reduced
by manipulating  phonon disorder and/or 
phonon-interface scattering (avoiding concomitantly  drastic changes in the electronic system). Our system is then expected to possess a high figure of merit.

\subsection{Three-terminal  figure of merit} 

\label{3tf}
The three-terminal geometry suggests novel possibilities for thermoelectric
applications. For example, when $\Delta T<0$ and $\delta \mu>0$, the
setup serves as a refrigerator of the local phonon
system, whose efficiency is given by the rate of the heat pumped from the phonon system to the electrical work invested, 
\begin{align}
\eta =I^{pe}_{Q}/(I^{}_{e}\delta\mu )\ .\label{FIRE}
\end{align}
Consider first
the special situation with $\ov{E}=0$, where
$L_1=K_e^0=L_{3}=0$. 
For a given $\Delta T$, $\delta\mu$ is adjusted to optimize the efficiency, yielding
\begin{align}
\eta=\eta_{0}(2+\tilde{Z}T-2\sqrt{\tilde{Z}T+1})/(\tilde{Z}T)\ ,
\end{align}
where $\eta_{0}=T/|\Delta T|$   is the Carnot 
efficiency, and the new figure of merit is
\begin{align}
\tilde{Z}T=L^{2}_{2}/(GK^{}_{pe}-L^{2}_{2})\ .
\end{align}
Inserting here Eqs. (\ref{TC}) yields  $\tilde{Z}T\rightarrow\infty$
upon neglecting the ``parasitic" conductances, discussed below.
When $\ov{E}\ne 0$, such an optimization can be
achieved by setting $\delta T=0$. 

In reality, $\tilde{Z}T$ must be finite. To calculate a more  realistic efficiency,
we generalize Eq. (\ref{FIRO})
by adding
the elastic transmission, the tunneling conductance $G_{el}$, to 
the hopping conductance $G$, and the elastic components, 
$L_{1,el}$ and $K_{e,el}^{0}$, to $L_{1}$ and $K_{e}^0$.
(The elastic
transmission does not contribute to $L_2$,  $L_{3}$,  and $K_{pe}$,
which are related to the heat transfer between the electronic and
phononic systems.\cite{3t})
We also include the phonon heat conductance $K_p$, replacing $I^{e}_{Q}$ by
$I_Q$,  the total heat current from the left to the right lead
($\delta T$ is now also the temperature difference for the phononic
systems in the left and right leads). 
Due to the absence of
phonon-drag effects in localized electronic systems, the
temperature difference $\delta T$ should not contribute to other
currents beside $I_Q$. Finally, there are phononic heat
 flows from the two leads to the system being cooled. Hence the numerator of 
 Eq. (\ref{FIRE}) is replaced by $I_Q^{pe}-K_{pp}\Delta
  T/T$, where $K_{pp}$ describes the phononic heat conductance in
such processes.

Following the same procedure as above, the efficiency is optimized by
adjusting $\delta \mu$ at $\delta T=0$. The result is similar, except
that the figure of merit is modified, 
\begin{align}
\tilde{Z}T&=\frac{L^{2}_{2}}{(G+G^{}_{el})(K^{}_{pp}+K^{}_{pe})-L^{2}_{2}}\nonumber\\
&
=\Bigl [\frac{G^{}_{el}}{G}+\frac{K^{}_{pp}}{K^{}_{pe}}+\frac{G^{}_{el}K^{}_{pp}}{GK^{}_{pe}}\Bigr ]^{-1}\ .
\end{align}
This  has a straightforward physical interpretation: The wasted work
is due to the elastic conductance and the unwanted heat diffusion, and 
$\tilde{Z}T$ is limited by the ratio of the waste to the useful powers.
In nanosystems $K_{pp}$ can be limited by the contact between the
system and the leads. Hence the ratios can be made small and
$\tilde{Z}T$ can still be {\em large}. The three-terminal
device can also serve as a heater and as a thermoelectric battery, where the same
figure of merit describes the efficiency.\cite{Honig}

\begin{figure}[htb]
\includegraphics[height=4.cm]{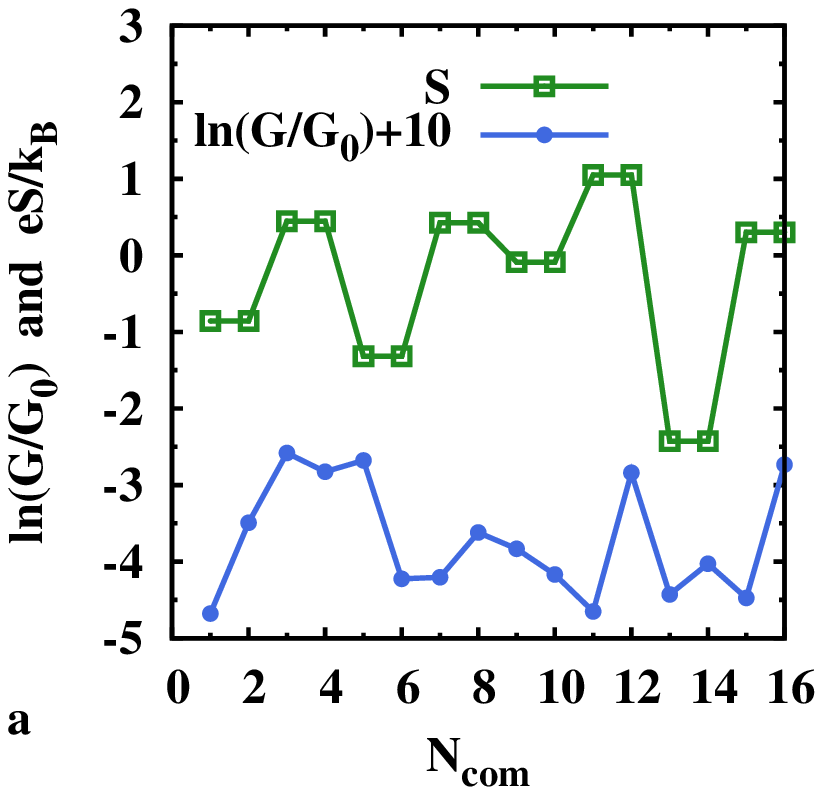}\includegraphics[height=4.cm]{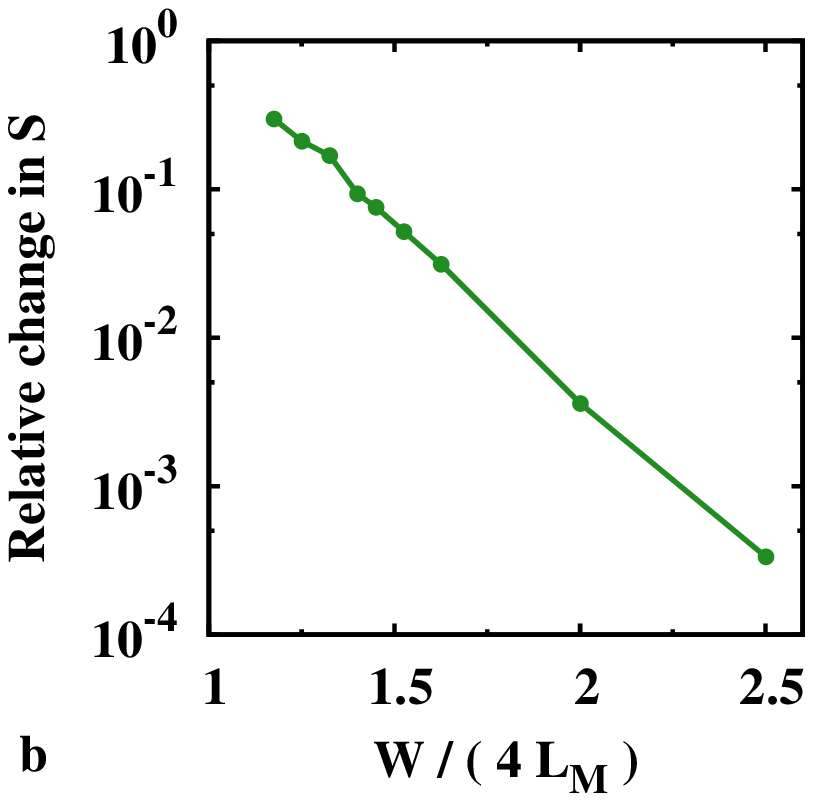}
\caption{(Color online) a. The conductance,  $\ln(G/G_0)$,  and thermopower, 
  $eS/k_{\rm B}$; the abscissa gives the number of computations, $N_{\rm
  com}$. The parameters are  $W=800, L_M=20, \xi=3.3$, in units 
average nearest-neighbor distance, $T_0=3000$, $T=20$, and the
energy-band  width is $1090$ ($T_0$ and $L_M$ are the Mott \cite{ZO}
temperature and  length, defined in the text).  b. The relative change in $S$ as a function of the
system  length $W/(4L_{\rm M})$, obtained by averaging over $10^6$
random configurations. Parameters (except $W$) are the same as in a.} 
\label{fig3}
\end{figure}

\subsection{Longer 1D systems}
For a chain of localized states, the picture is similar though
slightly more complex. Consider first nearest-neighbor hopping, where
the system is  a chain of resistors.  The same considerations as in
the two-site case  [in which the energy transferred is determined by  site 1
(2) and the left (right) lead] hold here for the leftmost, $\ell$,
(rightmost, $r$) localized state and the left (right) lead. Hence,  
the thermoelectric transport is  described by
Eqs.~(\ref{IQ})-(\ref{SP}), with $E_{1}$ ($E_{2}$) replaced by
$E_{\ell}$ ($E_{r}$). In particular, the thermopower coefficients $S$
and $S_{p}$ are {\em completely determined} at the left and right {\em
  boundaries}, despite the fact that  transport coefficients are
usually determined by both the boundaries and the  ``bulk".


In the  variable-range hopping regime, the result is similar:
$S$ and $S_{p}$ are determined by the resistors closed to the left and
right boundaries, within a distance comparable to the Mott length. 
This observation is confirmed by numerical simulations. In
Fig.~\ref{fig3}a we plot the thermopower $S$ and the conductance 
$G$ for different random configurations in which the energies $E_i$ and 
locations $x_i$ of the sites are random: The $E_i$ are chosen from a
uniform distribution in the range $[-E_{\rm max}, E_{\rm max}]$, with
$E_{\rm max}=545$   (units are defined in the figure caption) being larger than the hopping energy, $\sim
(T_0T)^{1/2}$,   determined by the Mott  \cite{vrh} temperature  $T_{0}$ which is of the order of the level spacing on scale $\xi$). The locations $x_i$ are chosen from a uniform distribution
in the range $[0,W]$, where $W$ is the length of the sample in units of the nearest-neighbor distance. The conductance of the whole network is
calculated by solving the Kirchhoff's equations,\cite{vrh} which yields
the currents through each bond for a given source-drain bias. The
thermopower $S$ is obtained from the particle current, $I_{N}$, and
the heat current $I_Q^e$ via  the relation $S=I_Q^e/(I_NeT)$.

The heat current $I_Q^e$ (particle current) is calculated by
summing over the heat currents (particle currents) flowing through  all
the bonds {\em connected with the leads}. In each such bond ($iL$)
[($iR$)] which connects the $i-$th localized state to the left (right)
lead, the heat current flow is $I_{Q}^{(iL)} =
E_{i}^{}I^{(iL)}_{N}$        [$I_{Q}^{(iR)} = E_{i}^{}I^{(iR)}_{N}$]
with $I^{(iL)}_{N}   $  [$I^{(iR)}_{N}   $] 
being the particle current in that bond. The total heat and particle
currents are $I^{e}_{Q}=0.5\sum_{i}(I^{(iL)}_{Q}+I^{(iR)}_{Q})$ and
$I_{N}=\sum_{i}I^{(iL)}_{N}=\sum_{i}I^{(iR)}_{N}$, respsectively.

At each $N_{\rm com}-$th computation, with $N_{\rm com}$ being an odd number,   a new random
resistor network is generated. At the subsequent, $N_{\rm com}+1-$th ,  computatio only the middle half part $[W/4,3W/4]$ of the network
is replaced by a new random configuration, while the parts close to the left and right boundaries, $[0,W/4]$ and $[3W/4,W]$,  are {\em not}
modified. It is seen from Fig. ~\ref{fig3}a that the conductance $G$
changes dramatically when the central part is modified, whereas the
thermopower is practically {\em immune to modifications of the central part}. To further study the sensitivity of the thermopower to
the sites that are a  distance larger than $W/4$ away from the two interfaces as a
function of $W$, we plot the relative change of the thermopower $S$  as a function of $W$ in Fig.~\ref{fig3}b. This relative change is defined as
$|S_{2n+1}-S_{2n+2}|/|S_{2n+1}+S_{2n+2}|$,  where
$S_{2n+1}$ and $S_{2n+2}$ denote the thermopowers calculated in the $2n+1$   and $2n+2$-th computation,  respectively. The results are obtained by averaging over $10^6$ random
configurations. It is seen that the relative change in $S$ decays
exponentially with increasing $W$, implying that sites located several Mott hopping distances $L_{M}$ away from the boundaries have a  
negligible effect on the thermopower $S$. The Mott length, $L_{M}$, is of the order of $[\xi/(\nu k_B T)]^ {1/(d+1)}$, where $\nu$ is the density of states at the Fermi level.

\vspace{0.8cm}

\section{Summary and Conclusions}
\label{SC}

In this work we studied the three-terminal thermoelectric transport
and thermopower mainly in simple 1D hopping systems in the linear-reponse regime. We worked
out the figure of merit for the three-terminal thermopower and
expressed it as a function of the three-terminal thermoelectric transport
coefficients. We obtained expressions for the thermoelectric transport coefficients in the simple two-site case. The system studied exhibits a large
thermopower and high figure of merit in the appropriate cases. We analyze the
conditions for high figure of merit in reality. 

For longer 1D chains, we found that, 
contrary to intuition based on the usual conductances, the thermoelectric transport coefficients in
hopping systems are {\em solely} determined by the states which are
close to the interfaces (approximately, within the relevant hopping
length). More details on these surprising results and their
generalizations will be given elsewhere. Finally, it should be
emphasized that {\em all} the results obtained here are in agreement
with the systematic microscopic derivations which will appear in consequent
work. Both Eq.~(\ref{Gtun}) and Eq.~(\ref{TC}) agree with the results derived
from the non-equilibrium Green function method.

\section*{Acknowledgments} We thank M. Pollak,  A. Rosch, 
P. W\"olfle and A. Amir for illuminating
discussions. OEW acknowledges the support of 
the Albert Einstein Minerva Center for Theoretical
Physics, Weizmann Institute of Science.
This work was supported by the BMBF within the
DIP program, BSF, by the ISF, and by its Converging
Technologies Program.

\end{document}